# Differential Conductance Fluctuation of Curved Nanographite Sheets in the Mesoscopic Regime

Haomin Wang,[1,2] Catherine Choong,[1,2] Jun Zhang,[2] Kie Leong Teo,[1] Yihong Wu[1,*]

[1] *Department of Electrical and Computer Engineering, National University of Singapore, 4 Engineering Drive 3, Singapore 117576*

[2] *Data Storage Institute, 5 Engineering Drive 1, Singapore 117608*

**Abstract**

Excess conductance fluctuations with peculiar temperature-dependence from 1.4 to 250 K were observed in curved nano-graphite sheets with an electrode gap length of 300 and 450 nm, whereas the conductance fluctuation is greatly suppressed above 4.2 K when the electrode gap length increases to 800 and 1000 nm. The former is discussed in the context of the presence of a small energy bandgap in the nano-graphite sheets, while the latter is attributed to the crossover from coherent transport to diffusive transport regime.

*PACS*: 73.23.-b, 73.63.-b, 73.40.–c, 81.05.Uw

*Keywords:* A. Nanographite sheets; D. Electron transport; Differential conductance fluctuations

---

[*] Corresponding author. Tel.: +65-65162139; fax:+ 65-6779-1103.

*Email address*: elewuyh@nus.edu.sg (Yihong Wu).

The interplay between disorder and quantum interference plays a crucial role in determining the characteristics of electron transport in metals in the mesoscopic regime [1]. In a weakly disordered system, quantum interference between self-returned and multiply scattered paths of electrons on the scale of phase coherence length, $L_\phi$, leads to quantum corrections to the electrical resistance, which manifests itself in the form of weak localization (WL) [2] and universal conductance fluctuations (UCF) [3]. The UCF is of universal nature because its magnitude, on the order of the conductance quanta $e^2/h$, depends on the shape of the conductor but not on its size or strength of disorder. In addition to UCF caused by internal scattering centers, the quantum interference among different transport channels or due to multiply reflected electron (hole) waves from the electrodes also causes fluctuations in the conductance. The latter should become more prominent in samples with reduced backscattering but a long coherent length. In this sense, nanostructured two-dimensional (2D) carbons, including both single and multiple layer graphene sheets, are of particular interest for studying mesoscopic transport because of their unique electronic band structures [4] and the associated robust transport properties [5] even in highly disordered samples. The band structure of graphene is characterized by linearly dispersed conduction and valence bands which touch each other at the K and K' points in the Brillouin zone [6]. For nano-sized graphene or so-called graphene nano-ribbons, theoretical studies have shown that an energy gap opens at the K and K` points with a bandgap being inversely proportional to the ribbon width [7]. In addition to the bandgap, a narrow and flat band is also predicted to exist at the middle of the bandgap, depending on the atomic configuration of the edges or extended defects [8]. It has been reported in literature that a small bandgap is also present in nano-sized bilayer or few layers of graphene (FLG) sheets [9]. From the point of view of both fundamental physics studies and potential applications, it is of importance to understand how this small bandgap would affect the transport properties of 2D nanocarbons in both the coherent and classic diffusive regimes. In the former case, of our particular interest are the structures in which (i) the bandgap is comparable to $E_\phi = \hbar D / L_\phi^2$ (or $E_B = \hbar v_F / L$ in the ballistic case) or k$_B$T and (ii) $L \leq \min(L_\phi, L_T)$, where $D$ is the electron diffusion coefficient, $L$ is the size of the sample, $L_T$ is the thermal length, $v_F$ is the Fermi velocity, $\hbar$ is the Plank constant and $k_B$ is the Boltzmann constant. It is worth pointing out that most of the mesoscopic transport studies on semiconductors carried out so far have been focused on systems with a bandgap which is much larger than the aforementioned energy scales. The small bandgap and long coherent length of FLG ribbons make them distinguished from other systems and serve as an excellent platform for coherent transport studies.

The samples under study are carbon nanowalls (CNWs) grown by microwave plasma enhanced chemical vapor deposition [10]. As we reported previously, the CNWs are curved FLG sheets which grow almost vertically on the substrate surface and form a mutually supported network structure. Fig. 1(a) shows a typical scanning electron micrograph of CNWs. The width of each piece of nanowall is about 0.5 – 1 μm and the thickness is around 1 – 10 nm. Both high resolution transmission electron micrograpy (HRTEM) and Raman studies have shown that the CNWs are nanographite sheets with a high degree of disordering and high density of defects [10]. The unique surface morphology of CNWs makes it difficult to form top electrodes with small distances. In order to overcome this difficulty, in this work, bottom electrodes were employed to form the electrical contacts to the nanowalls. The use of bottom electrodes has three advantages over the top ones: (i) it is possible to fabricate electrodes with a spacing down to deep sub-micron using e-beam lithography, (ii) the pre-cleaning in hydrogen plasma prior to the growth and simultaneous annealing during the growth make it possible to form a low resistance contact between CNWs and Ti (100 nm in thickness) and (iii) vertically aligned CNWs allows to form direct contact of all the layers with the electrodes. The patterned electrodes were prepared using the electron-beam lithography in combination with liftoff processes (Fig. 1(b)). The 4 branches of the electrodes were used for four-probe measurement of the transport properties. The distance between the electrodes was varied from 300 nm to 1 μm, while the width of the contact is fixed at 1 μm. Subsequently, the CNWs were deposited onto the substrates with the patterned electrodes. Although the CNWs grow non-selectively on the entire substrate, the measured properties are mainly due to the portions within the gap because of the low resistance of the electrodes. In order to have a rough idea on how many pieces of CNWs exist in the gap region between the electrodes, we superimpose the electrodes (dotted line) on the nanowalls in Fig. 1(a). As can be seen from this figure, there are at most 1 to 3 pieces of nanowalls in the gap region. The four-probe transport measurements were conducted in a liquid helium cryostat having a lowest temperature of ~1.4 K. A magnetic field of up to 6 Tesla can be applied to the sample. Differential conductance measurements were conducted using the combination of Keithley 6220 current source and 2182A nanovoltmeter. All the measurements were carried out after the temperature was stabilized at each setting point.

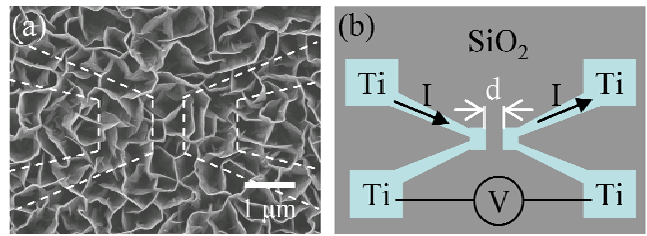

Fig. 1. (a) SEM image of CNWs. Dotted lines represent the electrodes configuration; (b) Schematic illustration of the electrodes before CNWs deposition. Current and voltage probes are indicated in the diagram. The electrodes are separated by a gap of "d" which varies between 300 nm and 1μm.

The transport properties of four Ti/CNWs/Ti samples with different electrode gaps (300 nm, 450 nm, 800 nm and 1μm) were measured by varying the temperature from 1.4 K to 250 K. As shown in Fig. 2(a), the temperature dependence of zero bias resistance (ZBR) shows semiconductor-like characteristic for all the samples between 1.4 K and 250 K. Similar type of behavior has been observed in single-walled carbon nanotube strands and was attributed to the existence of a super small bandgap in the specific samples [11]. As we mentioned in the introduction, recent theoretical calculations predict that a small energy gap should also be present in graphene nano-ribbons [7,12], and 2 to 3 layers of graphene sheets [9]. The latter has been observed experimentally [13]. Comparing to these perfect structures, the CNWs under study are more complex due to the presence of curvature and high degree of disorder and defects. Nevertheless, as revealed by HRTEM and Raman [10], the CNWs are still dominantly graphite in nature; thus we can presume that a small energy gap is also present in the CNWs. Depending on the atomic configuration of the edges and extended defects, there is also a possibility that a narrow and flat band may form at the middle of the bandgap [8]. Based on these assumptions, we adopted the simple two band (STB) model to analyze the experimentally observed temperature-dependence of resistance. The STB model has been often used to describe the electron transport in graphite [14]. Based on this model, the densities of electrons (n) and holes (p) are given by $n = C_n k_B T \ln(1 + \exp(-\frac{E_C - E_F}{k_B T}))$ and $p = C_p k_B T \ln(1 + \exp(-\frac{E_F - E_V}{k_B T}))$, respectively. Here, $E_F$, $E_C$ and $E_V$ are the energies at the Fermi level, bottom of conduction band and top of valance band, respectively, $k_B$ is the Boltzmann constant and $C_n$, $C_p$ are constants independent of temperature (T). Ignoring the contribution from static scattering centers, the mobility of carriers can be expressed as $\mu_e = A_1 T^{-1}$ or $\mu_h = A_2 T^{-1}$, where $A_1$ and $A_2$ are constants depending on the strength of electron and hole-phonon scattering in graphite. Since the resistivity is given by $\rho = (n\mu_n e + p\mu_p e)^{-1}$, the temperature dependence of resistance can be expressed as:

$$R = \frac{P_1}{\ln(1 + \exp(-\frac{E_C - E_F}{k_B T})) + P_2 \ln(1 + \exp(-\frac{E_F - E_V}{k_B T}))} + R_{contact}.$$

Using $E_C - E_F$, $E_F - E_V$, $P_1$, $P_2$ and $R_{contact}$ as the fitting parameters, this model fits well with our experimental data for the four samples, as shown in Fig. 2a. Based on this model, the energy gaps obtained for the four samples are 2.28±0.22 meV, 2.03±0.27 meV, 3.72±0.43 meV, and 1.62±0.09 meV for the samples with a gap of 1μm, 800 nm, 450 nm and 300 nm, respectively. These values are in good agreement with those found in literature [15], which presents an empirical relationship between the energy gap $E_g$ and width W in graphene ribbons: $E_g \approx 2eV \cdot nm/W$. If we use this formula to calculate the ribbon width based on the energy gaps obtained above, the width of the ribbon tuned out to be in the range of 0.5 – 1 μm. The height of carbon nanowalls is about 1-1.5 μm. As we are unable to control the orientation of the nanowalls with respect to the electrode edges, the actual length of CNWs inside the gap is not necessarily the same as that of the spacing between the electrodes. As far as quantum confinement is concerned, the energy gap is determined by both the height and width of the CNWs (of course, thickness may also play a role here). These factors may explain why a smaller energy gap was obtained for the sample with an electrode spacing of 300 nm as compared to those of the other three samples.

We now turn to the conductance fluctuations in mesoscopic CNWs. The underlying physics of UCF or quantum interference of multiply reflected electron waves is the difference in phase shifts experienced by electron waves that traversed a conductor along different trajectories within the phase coherence length. The phase of electron wave is sensitive to changes in configuration of scattering centers, magnetic field and electrostatic potential. The latter can arise from the modulation of chemical potential using a gate or application of a bias voltage in the electron traveling direction [16]. As the CNWs grow almost vertically on the substrate, the natural way to study both UCF and electron quantum interference due to scattering from the electrodes is to investigate how the conductance fluctuates with the DC bias rather than gate voltage or magnetic field. To this end, we have carried out detailed measurements of dI/dV curves at different temperatures for samples with different electrode gaps.

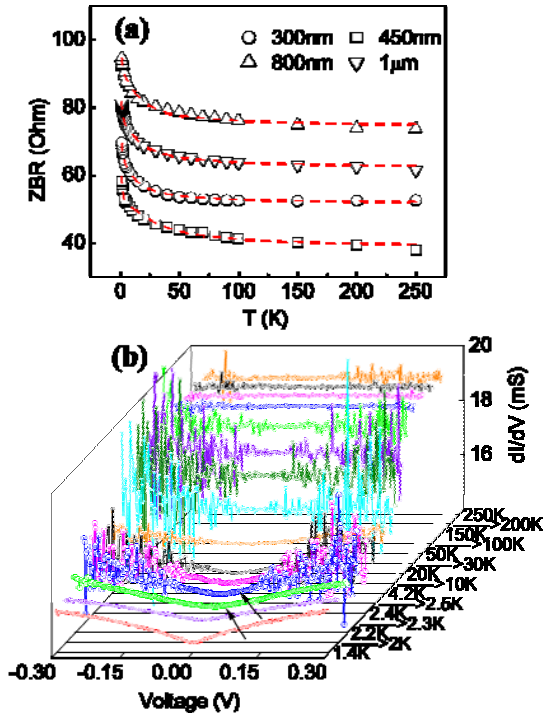

Fig. 2 (color online). (a) Plot of zero bias resistance versus temperature for three Ti / CNWs / Ti samples with an electrode spacing of 300 nm (circle), 450 nm (square), 800 nm (upward triangle) and 1 μm (downward triangle), respectively. Dashed-lines are fits with the STB model; (b) Differential conductance curves at temperatures from 1.4 K to 250 K plotted as a function of applied bias voltage V for the sample with an electrode spacing of 300 nm.

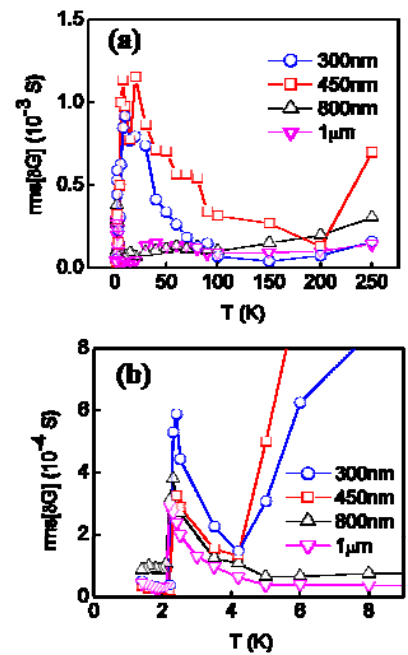

Fig. 3. (a) Plot of rms[δG] versus T for the three Ti / CNWs / Ti samples in the range of 1.4 K – 250 K; (b) Close-up of (a) in the low temperature region.

Some of the common features that have been observed include (i) the dI/dV is almost independent of the bias voltage above 250 K, and it evolves gradually into a V-shape when the temperature decreases to below 10 K, (ii) the conductance fluctuation is small at both high and very low temperature, but increases significantly in the temperature range between 2.2 K and 50 - 70 K, depending on the electrode spacing, (iii) the conductance fluctuation decreases significantly for the samples with electrode spacing of 800 nm and 1μm, and (iv) the conductance fluctuation increases with the bias. As can be seen from Fig. 2(b), the conductance fluctuations are sensitive to both the bias voltage and temperature. To characterize the temperature-dependence of conductance fluctuation quantitatively, we calculate the root mean square of the magnitude of dI/dV fluctuation $rms[\delta G]$ from − 5mA to 5mA at each temperature point as follows:

$$rms[\delta G](T) = \sqrt{\frac{1}{N}\sum_{i=1}^{N}(G_i(T,V_i) - \overline{G}(T,V_i))^2},$$

where $G_i(T,V_i)$ and $\overline{G}(T,V_i)$ are experimental and smoothened values of the differential conductance at temperature T and bias voltage $V_i$. The smoothened curve was approximated by a six-order polynomial trend line. The $rms[\delta G]$ versus temperature graph is plotted and shown in Fig. 3(a). The close-up of the low temperature region is shown in Fig. 3(b). The $rms[\delta G]$ shows a sharp peak at approximately 2.3 K for the 800 nm sample, 2.1 K for the 1 μm sample and at 2.2 K for the other two samples. When the temperature increases further, the $rms[\delta G]$ first decreases, reaches a minimum at about 4.2 K and then increases again for the samples with an electrode gap of both 300 nm and 450 nm. The second peak rises quickly at 9-10 K with a broad high temperature tail, centering at about 20 K and 50 K for samples with an electrode spacing of 300 nm and 500 nm, respectively. A further increase of temperature results in a sharp decrease of the conductance fluctuation at 50 K and 70 K, respectively, for the 300 nm and 450 nm samples. The former decreases to almost minimum at 100 K, while the latter decreases to the minimum at about 200 K. On the other hand, the $rms[\delta G]$ for the 800 nm and 1μm samples are small and almost constant above 4.2 K.

The appearance of conductance fluctuations and their unique temperature dependence can be understood by looking at the energy and length scales of electrons in this system, as discussed in the introduction. The conductance fluctuation becomes prominent in the conductance measurement when the electrode spacing becomes comparable to the phase coherence length or thermal length, whichever is smaller. By using the typical momentum relaxation time, $\tau = 1 \times 10^{-13} s$, and phase relaxation time, $\tau_\phi = 6.5 \times 10^{-12} s$ [17], for graphite sheets at 4.2 K and $v_F = 1 \times 10^6 m/s$, one has $D = v_F^2 \cdot \tau/2 = 0.5 m^2 s^{-1}$, $L_\phi = (D\tau_\phi)^{\frac{1}{2}} = 567 nm$, $L_T = 300 nm$, $E_\phi = 1.025 meV$ and $E_B = 2.2 - 6.6 meV$ (at $L = 300 - 100 nm$). This may explain why the magnitude of conductance fluctuations is large in the two samples with an electrode spacing of 300 nm and 450 nm, whereas it is small in the other two samples with a larger electrode gap above 4.2 K. We now look at the temperature-dependence of $rms[\delta G]$. The magnitude of UCF at zero temperature is of order of $e^2/h$, regardless of the sample size and degree of disorder. However, when temperature increases, the coherence length will decrease accordingly. If the coherence length is of the same order of sample size at zero temperature, the magnitude of UCF is expected to decrease monotonically with temperature in normal metals. The conductance fluctuation due to quantum interference of electron waves scattered from the electrodes is also expected to increase with decreasing the temperature, though the magnitude of fluctuation shall depend on the sample size and structure. However, the temperature-dependence of conductance fluctuations shown in Figs. 3(a) is far from this kind of "normal" behavior. Instead of monotonic decrease with temperature, the conductance fluctuations for samples with an electrode spacing of 300 nm and 450 nm exhibits two peaks, one at about 2.2 K for both samples and the other broad peak centering at ~ 20 K for the 300 nm sample and at ~ 50 K for the 450 nm sample. The latter agrees well with the bandgap of 1.62±0.09 meV and 3.72±0.43 meV for the samples with an electrode gap of 300 nm and 450 nm, respectively. This implies that thermal excitation of electrons from the valence band to the conduction band dominates the mesoscopic transport in CNWs above 4.2 K. In addition to global heating, the much enhanced fluctuations at high bias may also be due to current-induced local heating effect. It should be noted that the amplitude of conductance fluctuation in the peak temperature region is about one order larger than the UCF, in particular, at high bias. This again manifests the effect of thermal excitation across the small bandgap.

The minimum of conductance fluctuation appears around 4 K with amplitude on the order of $e^2/h$. This means that thermal-induced excitation is suppressed at low temperature. However, what is puzzling is the sharp upturn of conductance fluctuations below 4 K and an abrupt decrease again of conductance fluctuations below 2.1 - 2.2 K. A systematic study by using other different types of electrodes including superconductors revealed that this trend is independent of the electrodes and samples and was observed in all measurements. As the onset temperature of conductance fluctuation reduction is near the "lambda point" of helium 4 (He 4), we believe that the conductance fluctuation suppression below 2.1 K is presumably caused by the formation of a layer of superfluid He 4 on the nanowalls. The large thermal conductivity of superfluid He 4 greatly improves the temperature homogeneity of the nanowalls between the two electrodes, leading to the suppression of conductance fluctuation even at a high DC bias. By same reasoning, the upturn between 2.1-4 K can be easily understood as being caused by the thermal fluctuation due to the formation of piece-wise discontinuous regions of superfluid He 4.

In summary, we presented an electrical transport study of CNWs with low-resistance Ti electrodes. Excess conductance fluctuations were observed in the temperature range between 4 K and 200 K, which were attributed to the quantum interferences effect under the influence of thermally induced carrier excitation across a narrow bandgap. On the other hand, the sharp decrease of conductance fluctuation below 2.1 K is accounted for by the formation of a layer of He 4 superfluid on the nanowalls. The results obtained here have important implications for potential application of FLGs in electronics devices. We noted that, very recently, conductance fluctuations have also been observed in nano-ribbon based field effect transistors [15,18] and graphene [19] at low temperature. The conductance fluctuations found in graphene [20] is contributed to phase coherence multiple reflection of carriers between electrodes.